# A HIGH SPEED UNSUPERVISED SPEAKER RETRIEVAL USING VECTOR QUANTIZATION AND SECOND-ORDER STATISTICS

*Konstantin Biatov*

*konst.biatov@googlemail.com*

## ABSTRACT

This paper describes an effective unsupervised method for query-by-example speaker retrieval. We suppose that only one speaker is in each audio file or in audio segment. The audio data are modeled using a common universal codebook. The codebook is based on bag-of-frames (BOF). The features corresponding to the audio frames are extracted from all audio files. These features are grouped into clusters using the K-means algorithm. The individual audio files are modeled by the normalized distribution of the numbers of cluster bins corresponding to this file. In the first level the k-nearest to the query files are retrieved using vector space representation. In the second level the second-order statistical measure is applied to obtained k-nearest files to find the final result of the retrieval. The described method is evaluated on the subset of Ester corpus of French broadcast news.

## 1. INTRODUCTION

Currently a lot of audio resources are available. The web 2.0 platform gives possibilities to work online with the large amount of multimedia content. The information retrieval methods which can provide quick, reliable and efficient search in a large database are required. In the last decade some methods for audio query-by-example retrieval are suggested [1], [2], [3], [4]. In [1] for query-by-example audio search for similarity measure the authors suggest the technique based on Kolmogorov complexity. The retrieval is applied to four broad classes such as environmental sounds, music, singing and speech of 7 speakers. In [2] is considered the Euclidean distance between the GMMs of the query and audio database files. The distance measure between collection of distributions and its application to speaker recognition is described in [4]. The comparative study of speaker distinguishing distances is presented in [5]. The unsupervised query-by-example speaker retrieval task is closely related to the unsupervised speaker clustering. One of the initial work in an unsupervised speaker clustering is done in [6]. The extended review of an unsupervised speaker clustering and segmentation techniques are presented in [7]. The state-of-the-art technique for speaker clustering is based on the Bayesian Information Criterion (BIC). The BIC uses a second-order statistic. The BIC is successfully applied in initial [7] and in recent works for speaker clustering. The descriptions of some others second-order statistical measures used in the speaker recognition task are presented in [8], [9], [10]. The vector quantization technique for speaker recognition is still an active research area [11], [12], [13].

The presented paper proposes an unsupervised effective technique for speaker retrieval based on query-by-example. We apply the idea which was successfully used in speaker clustering task [14]. The speakers are modeled using a common universal codebook. The features corresponding to the audio frames are extracted from all audio files. These features are grouped into clusters using k-means algorithm. The individual audio files are modeled by the normalized distribution of the numbers of cluster bins corresponding to the feature-vectors of this file. The audio files are represented using a vector space model. For speaker retrieval a two-level method is used. In the first level using a vector space model representation the k-nearest to the query audio files are retrieved. Using cosine distance provides effective search in a large audio database. In the first level the number of k-nearest is relative large to capture all relevant speaker-files. On the second level more precise search is carried out. For retrieval the metric based on second-order statistic is used. Using a vector space model effectively provides reduced speaker search space and using second-order statistic measure produces precise and effective speaker retrieval in reduced search space. On both levels of retrieval different metrics are considered and compared.

## 2. SECOND-ORDER SRATISTICAL MEASURES

### 2.1. BIC based distance measure

The BIC for audio application was initially proposed in [6]. In general the BIC is defined as:

$$BIC(M) = \log L(X,M) - 0.5\,\lambda\ \#(M)\log(N,)\qquad(1)$$

where $\log L(X,M)$ denotes likelihood of segment $X$ given by the model $M$, $N$ is the number of feature vector in the data, $\#(M)$ is the number of free parameter in the model and $\lambda$ is a tuning parameter. For Gaussian distributions in order to estimate the data distribution turn point between two segments $i$ and $j$ that have $n_i$ and $n_j$ frames respectively, the $\Delta$BIC value is computed as:

$$\Delta BIC = 0.5 n_i \log|\Sigma_i| + 0.5 n_j |\Sigma_j| - 0.5 n_{ij} \log|\Sigma_{ij}| + \lambda P\,,\ (2)$$

where $n_{ij} = n_i + n_j$, $d$ is the dimension of the feature vector, $\Sigma_{ij}$ is the covariance matrix of the data points from two segments $i$ and $j$, $\Sigma_i$ is the covariance matrix of the data points from the segment $i$; $\Sigma_j$ is the covariance matrix of the data points from segment $j$ and $P$ is:

$$P = 0.5(d + 0.5 d(d+1))\log(n_i + n_j)\ .\qquad(3)$$

The $\Delta$BIC is the distance between two Gaussian models. The negative value of $\Delta$BIC indicates that two models fit to the data better than one common Gaussian model. The positive value of $\Delta$BIC indicates statistical similarity of compared models. The $\Delta$BIC can be considered as distance based on second-order statistic and can be applied for audio retrieval.

## 2.2. Divergence shape measure

Let us define the speaker covariance model C as

$$C_S = \frac{1}{N-1} \sum_{i=1}^{N} (x_i - \mu_s)(x_i - \mu_s)^t \qquad(4)$$

where N is the number of training feature vectors, x is the i training vector and is $\mu$ the mean of the training data. Suppose we are using a single covariance. The divergence shape is an information-theoretic measure between two Gaussian classes introduced as a measure of dissimilarity between tested and reference speaker [9]. An expression that depends only on the covariance matrixes is named Divergence shape (DS):

$$DS(C_1, C_2) = \frac{1}{2} tr[(C_1 - C_2)(C_2^{-1} - C_1^{-1})]\ ,\qquad(5)$$

where tr(.) is the trace operator and C is the covariance matrix.

## 2.3. Arithmetic harmonic sphericity measure

The arithmetic harmonic sphericity (AHS) is defined in [8]. The AHS is computed as follows:

$$AHS(C_1, C_2) = \frac{1}{2} tr[(C_1 C_2^{-1})(C_2 C_1^{-1})]\qquad(6)$$

## 2.4. Hotelling $T^2$ statistic

The statistical test for significance of difference in mean vectors $\mu_1$ and $\mu_2$ between two sets of vectors is the Hotelling $T^2$ statistic. The common covariance matrix is estimated as:

$$C = ((N-1)C_1 + (M-1)C_2)/(N+M-2)\qquad(7)$$

Where N, $C_1$ number of vectors and covariance matrix of the first data set and M, $C_2$ number of vectors and covariance matrix of the second data set. The $T^2$ statistic is defined as:

$$T^2(\mu_1, \mu_2) = \frac{NM}{N+M}(\mu_1 - \mu_2)^t C^{-1}(\mu_1 - \mu_2)\qquad(8)$$

## 3. DISTANCE METRIC IN VECTOR SPACE REPRECENTATION

In the vector space model representation used in the first stage of the retrieval the following similarity measures are compared: the normalized L2 distance, the cosine distance, the Bhattacharyya distance and the distance based on histogram intersection. Let us denote the normalized histogram for file i as $f_{ij}, j \in \{1,...,M\}$, where M the number of bin in the codebook. The $|f_k|$ denotes the norm of the vector.

The cosine distance is defined as

$$dist(f_i, f_k) = \frac{\sum_j(f_{ij} f_{kj})}{|f_i||f_k|},\qquad(9)$$

the normalized L2 distance is defined as

$$dist(f_i, f_k) = \sum_j \left(\frac{f_{ij}}{|f_i|} - \frac{f_{kj}}{|f_k|}\right)^2\qquad(10)$$

the Bhattacharyya distance is defined as

$$dist(f_i, f_k) = \sum_j \sqrt{f_{ij} f_{kj}}\qquad(11)$$

the histogram intersection distance is defined as

$$dist(f_i, f_k) = \frac{\sum_j \min(f_{ij} f_{kj})}{\min(|f_i|, |f_k|)}\ .\qquad(12)$$

## 4. ALGORITHM DESCRIPTION

In the presented algorithm the speakers are modeled using a common universal codebook. The codebook is based on BOF. The features corresponding to the audio frames are extracted from all audio files. These features are grouped into clusters using k-means algorithm. The individual audio files are modeled by the normalized distribution of the numbers of cluster bins corresponding to the feature-vectors of this file. The audio files are represented using a vector space model. For speaker retrieval a two-level method is exploited. In the first level using a vector space model representation the k-nearest to the query audio files are retrieved. In this level the effective search in a large audio database could be provided. The number of k-nearest is relatively large to capture all relevant speaker-files. On the second level more precise search using second-order statistical measure is carried out. Using vector space model representation effectively reduces speaker search space and using second-order statistical measure produces final precise and effective unsupervised speaker retrieval in large audio database. On both levels of retrieval different metrics are considered and compared.

## 5. DATABASE DESCRIPTION

The data used in the experiments are from ELRA [15]. The collection is the corpus collected in the French national project ESTER (Evaluation of Broadcast News enriched transcription systems) [16]. The original transcriptions contain a total of 2,172 different speakers. About one third (744) are female speakers while 1,398 are male speakers. The data obtained from ELRA [15] includes 517 speaker and 47880 audio segments. The total duration is approximately 40 hours. In the experiments 7874 audio segments belonging to 300 speakers including 220 male and 80 female speakers are used. The speakers with music background which are from a commercial part are not included in the test data. Part of the files of the test set includes background noise. The narrow band speech is presented in 18% of the audio corpus. The duration of test data is 7.9 hours. The data is converted into 16kHz sampling rate and 16 bit per second. The audio signals are divided into overlapping frames with the size 25 ms. An overlap is 10 ms. The features consist of all-purposes MFCC. In the experiments we use 13 MFCC plus their first derivatives.

## 6. EXPERIMENTAL RESULTS

The performance of the retrieval experiments is measured in terms of percents of retrieving the relevant files in 1-best, 3-best and 5-best. For each segment available in database we retrieve 1, 3 or 5 nearest. In the experiments for audio retrieval **7874** audio segments belonging to **300** speakers with the total duration **7.9** hours are used. After feature extraction the frames with low energy are removed using threshold based classifier to avoid mismatch in audio files comparison because of the background noise. The distribution of segments duration of the test data after removing low energy frames is presented in Table 1.

| 1-2 seconds | 2-3 seconds | 3-4 seconds | 4-5 seconds | more 5 seconds |
|---|---|---|---|---|
| 0.012 | 0.370 | 0.308 | 0.193 | 0.116 |

**Table 1.** Distribution of the segments duration

For obtaining centroids for vector quantization and for reducing time of calculations we extract only 100 frames from each audio segment used in the test. Totally 787400 frames are extracted. The vector quantization is performed using a k-means algorithm with the 2048 centroids for BOF. The number 2048 is selected because the further increasing the number of centroids don't give significant improvement in the retrieval. In the first level of retrieval using vector space representation 50 nearest speaker-files is retrieved. The number 50 was selected experimentally based on retrieval performance and speed of the calculations. The performance of retrieval is evaluated for cosine distance, normalized L2 distance, Bhattacharyya distance and histogram intersection distance.
.

| Distance metric | 50 best |
|---|---|
| Cosine distance | 96.7% |
| Normalized L2 distance | 80.0% |
| Bhattacharyya distance | 96.9% |
| **Histogram intersection distance** | **97.6%** |

**Table 2.** The first level retrieval performance

The results of experiments show that histogram intersection distance has the best performance. In the retrieval on the second level four second-order statistical measures are compared: the distance based on BIC, the distance based on DS, the distance based on $T^2$ statistic and the distance based on AHS. The optimal value of $\lambda$ in BIC metric was 1.0. The results of experiments are presented in the Table 3. The distance based on BIC is performed as the best. The full retrieval time is also presented in the Table 3. The reduced search space obtained on the first stage provides the effectiveness and high performance on the second stage. The retrieval using only second-order statistic measure, for example BIC, has quadratic computational complexity with respect to the number segments and has limit of its application in large-scale data sets. The experiments are carried out on a computer with processor Intel (R), Core™ Quad CPU Q8300, 2.50 GHz.

| Distance metric | 1 best | 3 best | 5 best | Full time of retrieval |
|---|---|---|---|---|
| $T^2$ statistic | 30.9% | 57.1% | 69.7% | **1012 sec.** |
| DS | 69.9% | 86.2% | 90.9% | 1040 sec. |
| AHS | 69.3% | 85..8% | 90.7% | 1013 sec. |
| BIC | **72.6%** | **88.1%** | **92.1%** | 1319 sec. |

**Table 3.** The final level retrieval performance

## 9. CONCLUSIONS

The paper describes the method for speaker retrieval in audio database. For the retrieval the combination of some techniques is used. To obtain an effective and reliable search we suggest using a two-level retrieval technique. On the first level the effective search based on vector quantization and vector space representation is used. The result is 50 closest to the query audio files. For the first level retrieval 4 distance metrics are compared. The histogram intersection distance is the most reliable. The search on the first level reduces the search space for the second retrieval pass. Then for retrieval the distance measure based on second-order statistic is used. The retrieval is applied only to 50 nearest files retrieved on the first level. The BIC performs the best. It is more effective to provide search in restricted space than to compare all data. It requires the quadratic computational complexity with respect to the number of segments. The algorithm demonstrates high retrieval performance. The speed of the retrieval is 20 times more quicker than the duration of the processed data.

## 11. REFERENCES

[1] M. Helen and T. Virtanen, "A Similarity Measure for Audio Query by Example Based Perceptual Coding and Compression", in *Proceedings of the 10th International Conference on Digital Audio Effects (DAFx-07)*, Bordeaux, France, 2007.

[2] M. Helen and T. Virtanen, "Query by Example of Audio Signal Using Euclidean Distance between Gaussian Mixture Models", in *Proceedings of the IEEE International Conference on Acoustic, Speech and Signal Processing (ICASSP)*, Hawaii, USA, 2007.

[3] J. Jensen, D. Ellis, M. Christensen and S. Jensen, "Evaluation of Distance Measure Between Gaussian Mixture Models of MFCCs", in *Proceedings of 8th International Conference on Music Information Retrieval*, Vienna, Austria, 2007.

[4] H. Beigi, S. Maes and J. Sorensen, "A Distance Measure Between Collection of Distributions and its Application to Speaker Recognition", in *Proceedings of the IEEE International Conference on Acoustic, Speech and Signal Processing (ICASSP)*, Seattle, USA, 1998.

[5] A. Iyer, U. Ofoegbu, R. Yantorno and B. Smolenski, "Speaker distinguishing distances: a comparative study", Int. J. Speech Technology, (2007) 10: pp. 95-107.

[6] S.Chen S. and P. Gopalakrishnan, "Clustering via the Bayesian Information Criterion with the applications in speech recognition", *Proc. ICASSP'98*, 1998.

[7] M. Kotti, V. Moschou and C. Kotropoulos, "Review: Speaker segmentation and clustering", Signal Processing 88 (2008), pp. 1091-1124.

[8] F. Bimbot, I. Magrin-Chagnolleau, L. Mathan, "Second-Order Statistical Measure for Text-Independent Speaker Verification", *Speech Communication, Vol. 17, pp. 177-192*.

[9] J. P. Campbell, "Speaker Recognition: A Tutorial", in *Proceeding of IEEE*, Vol. 85, No. 9, pp. 1437-1462.

[10] R. Zilca, "Using Second Order Statistics for Text Independent Speaker Verification", In *Proceedings of 2001: A Speaker Odyssey – The Speaker Recognition Workshop*, June 2001, Crete, Greece.

[11] S. Furui, "Vector-Quantization Based Speech Recognition and Speaker Recognition Technologies", Signal, Systems and Computers, 1991. 1991 Conferecne Record of the Twenty-Fifth Conference on, Vol. 2, pp. 954-958.

[12] J. Pelecanos, S. Myers, S. Sridharan and V. Chandran, "Vector Quantization based Gaussian Modelling for Speaker Verification", in *Proceedings of the International Conference on Pattern Recognition (ICPR'00)*, Barcelona, Spain.

[13] C. Wang, Z. Miao and X. Meng, "Differential MFCC and Vector Quantization Used for Real-Time Speaker Recognition System", in *Proceedings of the 2008 Congress on Image and Signal Processing*, Vol. 5, pp. 319-323.

[14] K. Biatov, "A Fast Speaker Clustering Using Vector Quantization and Second Order Statistics with Adaptive Threshold Computation", in *Proceedings of INTERSPEECH 2010*, Japan.

[15] www.elra.info

[16] S. Galliano, E. Geoffrois, G. Gravier, J.-F. Bonastre, D. Mosrefa, K. Choukri, "Corpus description of the Ester Evaluation Campaign for the Rich Transcription of French Broadcast News", in *Proceedings of the 5th International Conference on language resource and Evaluation (LREC06)*, May 2006, Genova, Italy.